\documentclass[journal,article,accept,moreauthors,pdftex]{Definitions/mdpi} 
\usepackage{float}
\usepackage[utf8]{inputenc}
\usepackage[english]{babel}

\usepackage{natbib}

\firstpage{1} 
\pubvolume{xx}
\issuenum{1}
\articlenumber{5}
\pubyear{2019}
\copyrightyear{2019}
\history{Received: 13 April 2020; Accepted: 29 April 2020; Published: 6 May 2020}

\Title{An Intelligent Non-Invasive Real Time Human Activity Recognition System for Next-Generation Healthcare}

\Author{William Taylor $^{1}$*,  Syed Aziz Shah $^{2,}$ $^{1}$,  Kia Dashtipour $^{1,}$,  Adnan Zahid $^{1,}$,  Qammer H. Abbasi $^{1}$ and Muhammad Ali Imran $^{1}$}

\AuthorNames{William Taylor, Syed Aziz Shah, Kia Dashtipour, Adnan Zahid, Qammer H. Abbasi and Muhammad A. Imran}

\address{%
$^{1}$ \quad James Watt School of Engineering, University of Glasgow, Glasgow,G12 8QQ, U.K; kia.dashtipour@glasgow.ac.uk,a.azahid.1@glasgow.ac.uk,Qammer.Abbasi@glasgow.ac.uk, Muhammad.Imran@glasgow.ac.uk\\
$^{2}$ \quad School of Computing and Mathematics, Manchester
Metropolitan University U.K; S.Shah@mmu.ac.uk}

\corres{Correspondence: 2536400t@student.gla.ac.uk (W.T.)}

\abstract{Human motion detection is getting considerable attention in the field of Artificial Intelligence (AI) driven healthcare systems. Human motion can be used to provide remote healthcare solutions for vulnerable people by identifying particular movements such as falls, gait and breathing disorders. This can allow people to live more independent lifestyles and still have the safety of being monitored if more direct care is needed. At present wearable devices can provide real time monitoring by deploying equipment on a person's body. However, putting devices on a person's body all the time make it uncomfortable and the elderly tends to forget it to wear as well in addition to the insecurity of being tracked all the time. This paper demonstrates how human motions can be detected in quasi-real-time scenario using a non-invasive method. Patterns in the wireless signals presents particular human body motions as each movement induces a unique change in the wireless medium. These changes can be used to identify particular body motions. This work produces a dataset that contains patterns of radio wave signals obtained using software defined radios (SDRs) to establish if a subject is standing up or sitting down as a test case. The dataset was used to create a machine learning model, which was used in a developed application to provide a quasi-real-time classification of standing or sitting state. The machine learning model was able to achieve 96.70 \% accuracy using the Random Forest algorithm using 10 fold cross validation. A benchmark dataset of wearable devices was compared to the proposed dataset and results showed the proposed dataset to have similar accuracy of nearly 90 \%. The machine learning models developed in this paper are tested for two activities but the developed  system is designed and applicable for detecting and differentiating  x number of activities.}


\begin{document}


\section{Introduction}
Human motion detection is an important area of research in the field of healthcare systems. Eventually, more and more sectors of the healthcare industry will begin to use technology \cite{yang2020human,abbasi2016advances}. In recent years, home healthcare through the use of different technologies has gained much attention from its ability to improve the lives of people who require special care \cite{dong2017monitoring,al2018critical}. Special care is required by a large number of people such as the elderly population. The elderly population is on the rise, leading to a substantial decline in nursing home capacity. \cite{mercuri2016healthcare,haider2019efficient}. The elderly population is set to be 2.1 billion in the year 2050 according to statistics from the United Nations \cite{liu2019novel,fan2018breathing}. With this expected growth in elderly population, it will have even more strain on the lack of care givers, so that dependencies on the technology will be required to support the treatment \cite{shang2019bia}. Monitoring of elderly and vulnerable people can allow for them to live more independently. This means that the level of care they receive can be less. This is because the monitoring can provide real-time messages to care givers in the instance of a fall \cite{yang2018monitoring}. Human motion detection is the process of using technology to extract the features of the human movement \cite{zhang2019wigrus,tahir2019wifreeze,liu2018respiration}. Human motion detection can be used for the monitoring of patients and vulnerable people such as the elderly or young children \cite{yang2020diagnosis,demir2016anatomical}. Fall detection is just one example of how human motion can be used in the healthcare industry although an important example. The world health organisation reports that falls can cause around 646 thousand death and over 37 million serious injuries. \cite{santos2019accelerometer,jilani2018millimeter}. If a system was able to provide careers with this information in real time then the patient would be able to receive assistance from the carer without the carer having to be with the vulnerable person at all times contributing to a more independent lifestyle. Human movement can be detected by the use of wearable devices such as mobile or smart watches using accelerometers, which can then pass the information to carers or physicians etc. \cite{yao2019energy,yang2011spatial}. There leaves an issue of when the patient forgets to wear the wearable device. Another method of human motion detection is to use radio waves already in the atmosphere such as Wi-Fi in a home network. This technique is considered as non-invasive. Non-invasive is defined in medical terms as not involving the introduction of instruments into the body such as the case with wearable devices. This can be achieved by using the Channel State Information(CSI) from Wi-Fi to look at the amplitude of the CSI as a human moves between the radio waves \cite{zhao2019r,chopra2016thz}. The CSI is a feature in Wi-Fi that describes how the wireless signal propagates between the transmitting node and receiving node \cite{lolla2019wifi}. This data can be exploited to detected changes during a specific human motion. This research will explore the use of Universal Software-defined Radio Peripheral (USRP) to build a dataset of the CSI information of human activities and then use machine learning for binary classification of a human either sitting down or standing up. USRPs will be used because they offer a simple framework for experimentation rather than setting up complex systems for functionality testing \cite{christiansen2019development,demir2016anatomical}. USRPs are widely used in research applications because of their ability to transfer and receive frequencies in several bands \cite{kim2017device}. URSPs provide flexibility as they can be tuned to a wide range of frequencies \cite{tichy2012ofdm}. This work will use 64 subcarriers. Orthogonal frequency division multiplexing (OFDM) is used for 64 points of fast Fourier transformer (FFT) producing 64 frequency carriers (subcarriers) \cite{ashleibta2020software}. Lower frequencies are able to detect the smaller movements while higher frequencies are able to detect larger movements \cite{zhang2017wicare}. Using USRPs allow for a range of frequencies to be used in the experimentation which will allow a greater detection in movements overall. This paper aims to research the abilities of using RF signals to be able to classify human motion in a real time application. This paper reports two major contributions to the state of the art. The first contribution is presenting a simple set up of how a machine learning model can provide real time classification on human motion using data retrieved from a URSP. The second contribution is providing a comparison between the newly acquired dataset and an existing wearable device human motion dataset. This paper is organised in the following sections. Section 2 will detail some of the related work. Section 3 will detail the methods employed to collect the data. Section 4 will describe the methods of machine learning used and section 5 will display the results and discuss said results as well as compare the results to a benchmark dataset collected from wearable devices.
 
\section{Related Work}
This section looks at the recent literature in various forms of human motion detection and where machine learning has been applied. The articles in \cite{chin2019daily,shah2018seizure,fioranelli2019radar} collected a range of human activities where the test subjects were using wearable accelerometer on their wrists. The dataset collected by these activities were then run through the machine learning algorithms Random Forest, K Nearest Neighbours (KNN) and Support Vector Machine. The results found that Support Vector Machine had the highest results of 91.5 \%. The work of \cite{ding2019fall,shah2019human,shah2019rf} used frequency-modulated continuous-wave (FMCW) radar system to look at the Doppler, temporal changes and radar cross sections to collect data of falling and other fall related activities such as stepping, jumping, squatting, walking and jogging from 3 participants. The data was then run through 10 cross fold validation with KNN to achieve a high accuracy result of 95.5 \%. This work demonstrates that wireless waves can be used to classify human motion through the changes in frequencies. 
A similar work was done on multi channel extraction in \cite{liu2019noma,liu2018novel}. Jalal et al. \cite{jalal2019wrist} used a benchmark dataset of 14 indoor human activities. The benchmark dataset was collected using triaxial accelerometer sensors. The research included separating the static activities from the dynamic activities. The paper then went on to apply the random forest algorithm for machine learning classification. The static results scored higher at 92.16 \% with the dynamic activities scoring 80.0 \% with an average result of 85.17 \%. The work conducted in \cite{zhang2019recognizing} used wearable smart watches to monitor the movement of ping-pong players. The watch recorded data of 8 different motions on how the test subjects moved the ping-pong paddle including forehand attack, forehand flick, backhand flick etc. The data was then processed using 7 machine learning algorithms including Random Forest, SVM, KNN and decision trees. The research found Random Forest to be the best performance with an accuracy score of 97.80  The paper \cite{zhang2020complex} made use of CSI on Wi-Fi OFDM signals for the classification of 5 different arm movements. The human made different arm movements while standing between a Wi-Fi router and a laptop sending wireless signals to each other. The CSI was then captured and machine learning was applied to the collected data. The machine learning algorithm chosen was the Long Short-Term Memory (LSTM) which was able to achieve a high accuracy result of 96 \%. A similar work on healthcare was done in \cite{alintelligent,oueida2018edge,anjomshoa2017social}. Nipu et al. \cite{nipu2018human} used CSI information to try and identify a specific person. The experiment conducted had different people walk through two devices while data is transmitted and store the CSI information obtained while that person walked through the radio frequencies.  The dataset was then passed through the machine learning algorithms, Random forest and Decision tree. The experiments found that the algorithms scored higher when only 2 people were used in a binary classification experiment.

\section{COLLECTION OF DATA}
\begin{figure}[H]
    \centering
    \includegraphics[width=0.6\textwidth]{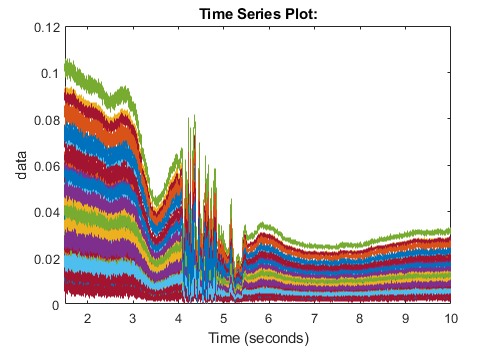}
    \caption{Channel State Information for the human motion of sitting down}
    \label{fig1}
\end{figure}

\begin{figure}[H]
    \centering
    \includegraphics[width=0.6\textwidth]{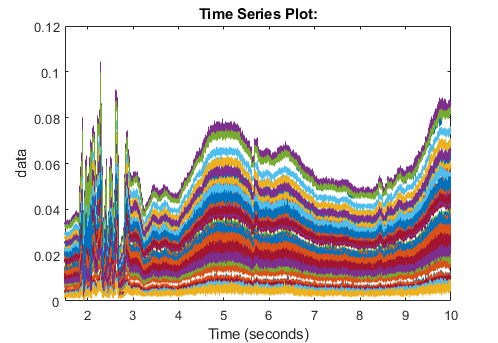}
    \caption{Channel State Information for the human motion of standing up}
    \label{fig2}
\end{figure}

In this section we will discuss the methods of how the data is collected. The work of this paper makes use of Universal Software Radio Peripheral (USRP) devices to send packets between antennas\cite{tanoli2018experimental}. Two USRPs were used namely the X310/X300 models from a national instrument (NI), each equipped with extended bandwidth daughterboard slots covering DC– 6 GHz with up to 120 MHz of baseband bandwidth. The X300 model was used as the transmitter with the X310 model performing as the receiver. The devices were connected to two PCs through 1G Ethernet cable connections. The USRP’s were equipped with of two VERT2450 omni directional antennas. The simulation was designed using MATLAB/Simulink program linked to the USRP’s. The experiment was undertaken in an office environment and USRPs were kept at 4 metres within line of sight with each other, to achieve the best performance. 
Experiments were performed with set parameters. Table \ref{tab:title} lists the parameters of the software configuration of the USPRs. The USRPs used in the study have a frequency range from 1 GHz to 10 GHz. Centre frequency for the USRPs was set as 5.32 GHz and the operational frequency of omni directional antenna was also 5.32 GHz, with 3 dBi gain. The gain of USRP chosen to be 70 for transmitter and 50 for the receiver. The hardware parameters values of the USRP is summarised in table \ref{tab:title1}. Ethical approvals of participants have been acquired through university of Glasgow ethic review committee. The participants were asked to perform the different human motions in this research of standing up and sitting down. Participants completed the task multiple times to be able to collect many samples of the CSI information to allow for error and allow cleanest samples to be taken forward. The test was performed in a 7 by 8 meters office space containing furniture such as tables, chairs, draws, etc. The human motion is then carried out between the antennas and the Channel State Information is then recorded while this human motion is carried out. As radio signal propagation is proportionate to the movement of the human, the CSI will differentiate as different motion takes place. The CSI will show certain properties when a certain movement is made by the human. In this paper we have recorded the CSI for multiple subjects sitting down on a chair and then standing from a chair. As there are many variations in the way the signals propagate and human movement will never be exactly the same, the movement should follow the same patterns in the CSI data. Some samples can be considered as good samples where interference is set to a minimum and some samples may be affected by ambient movement or atmosphere factors. Multiple samples are taken to try to capture the flow of the patterns and machine learning is used to attempt to classify the samples. The final dataset contains 30 samples each of sitting and standing. Figures \ref{fig1} and \ref{fig2} display the CSI of the 64 subcarriers of the USRP. Each colour represents a subcarrrier and the frequency of the subcarrier is shown along the Y-axis and time is shown along the X-axis while an activity is taking place. Figure \ref{fig1} shows the pattern followed in a good sample of sitting down and Figure \ref{fig2} shows the pattern followed in a good sample of standing up.

\begin{table}[H]
\caption {SOFTWARE CONFIGURATION PARAMETERS SELECTION} \label{tab:title} 
\centering
\begin{tabular}{||p{4cm} p{4cm} ||}  
 \hline
 Parameters & Values  \\ [0.5ex] 
 \hline\hline
 Input data (Signal) & round(0.75*rand(104,1))\\ 
 \hline
 Sample time & 1/80e4  \\
 \hline
 Modulation type & QPSK \\
 \hline
 Bit per symbol M & 2 bits \\ [1ex] 
 \hline
 OFDM Subcarrier & 64 subcarriers \\ [1ex] 
 \hline
 Pilot subcarrier & 4 \\ [1ex] 
 \hline
 Null subcarrier & 12 \\ [1ex] 
 \hline
 Cycle prefix M & NFFT-data subcarrier \\ [1ex] 
 \hline
 Samples per frame & Used subcarrier log2 (M) \\ [1ex]
 \hline
 \end{tabular}
\end{table}

\begin{table}[H]
\caption {HARDWARE CONFIGURATION PARAMETERS SELECTION} \label{tab:title1} 
\centering

\begin{tabular}{||p{4cm} p{4cm} ||}  
 \hline
 Parameters & Values  \\ [0.5ex] 
 \hline\hline
 Platform & USRP X300/X310\\ 
 \hline
TX IP address & 192.168.11.1  \\
 \hline
 RX IP address & 192.168.10.1 \\
 \hline
 Channel mapping & 1 TX, 2 RX \\ [1ex] 
 \hline
 Centre frequency & 5.32 GHz \\ [1ex] 
 \hline
 Local oscillator offset & Dialog \\ [1ex] 
 \hline
 PPS source & Internal \\ [1ex] 
 \hline
 Clock source & Internal \\ [1ex] 
 \hline
 Master clock rate & 120 MHz \\ [1ex] 
 \hline
 Transport data type & Int16 \\ [1ex] 
 \hline
 Gain (dB) & TX 70, RX 50 \\ [1ex] 
 \hline
 Sample time & 1/80e4 \\ [1ex] 
 \hline
 Interpolation factor & 500 \\ [1ex] 
 \hline
 Decimation factor & 500 \\ [1ex]
 \hline
 \end{tabular}
\end{table}

The USRPs are configured to transmit data from one antenna to the other for 10 seconds. As the signals propagate in different ways each time a sample is taken then the amount of packets received have slight variations. However this has little effect as the aim is to detect patterns in the radio waves as a certain human motion is carried out during the transmission of packets. 
Figure \ref{fig22} details the process used in this experimentation.

\begin{figure}[H]
    \centering
    \includegraphics[width=0.5\textwidth]{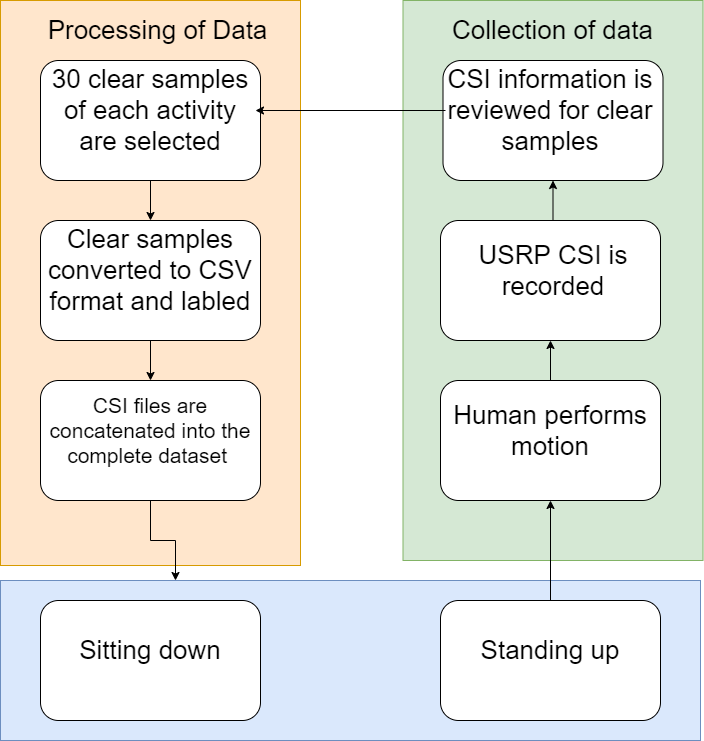}
    \caption{Experiment flow chart}
    \label{fig22}
\end{figure}

\section{Machine Learning Process}
The dataset performance has been measured using a range of machine learning algorithms using the Python SciKit library. Scikit is a machine learning package that is widely used in the data science field \cite{hao2019machine}. The Samples are converted into CSV format so that they can be processed using the SciKit library. The Python library Pandas is used to process the CSV files. Pandas imports the CSV files as dataframes within python which the SciKit library then processes \cite{pappalardo2019scikit}. The labels are added as the first column on the dataframes as the data is of varying length throughout the samples. Then the dataframe of each sample is combined together to make the full dataset, the varying lengths result in NAN values being part of the dataset. To resolve this issue SciKit provides a function called simple imputer. This is used to replace all NAN values with a 0. 
Therefore the shorter samples of the dataset will contain 0 values tailing the row on the CSV file. This is not perceived to be a problem as the differing lengths is minor and the pattern of the RF signals are still apparent. This is part of variance between different samples. The data set is then divided into two variables, one for the labels and one for the data itself. Then the four machine learning algorithms are declared. The four algorithms used to test this dataset are Random Forest, K nearest Neighbours, Support Vector Machine and Neural Networks. The ensemble classifier takes each algorithms prediction as a vote and then whichever prediction has the most votes will be the prediction declared by the ensemble classifier. 
\\
Random forest is a collection of decision trees. Each tree makes a prediction of the output by taking in looking for features found in the training phase. This prediction is considered a vote. The majority of prediction is the final Random Forest prediction \cite{shaikhina2019decision}. Equation \ref{eqn1} shows how SciKit uses Random Forest:

\begin{equation}
    \label{eqn1}
    Ni_j = W_jC_j - W_{left(j)}C_{left(j)} -W_{right(j)}C_{right(j)}
 \end{equation}
    
    \begin{itemize}
    \item $Ni_j $= the importance of node j
    \item $W_j $= weighted number of samples reaching node j
    \item $C_j $= the impurity value of node j
    \item $left_j $= child node from left split on node j
    \item $right_j $= child node from right split on node j
    \end{itemize}

The K nearest Neighbours algorithm is known for its simplicity. The algorithm works by comparing the testing data to the training data \cite{saccli2019microwave}. The features of the training data are assigned a K sample then the testing data is assigned to the K sample that nearest matches the new data \cite{li2019research}. Equation \ref{eqn2} shows the Euclidean KNN equation which is the default method for SciKit:
    \begin{equation}
    \label{eqn2}
    \sqrt {\sum_{i=1}^{k}(x_i - y_i)^2}
    \end{equation}
    
    \begin{itemize}
    \item $ {k} $ = is the number of samples
    \item $ {x} $ = the data
    \item $ {y} $ = the label
    \end{itemize}
    
The Support Vector Machine algorithm works by constructing hyper planes and uses these hyper planes to separate the input data into different categories. The training data is used to train the hyper planes based on features of the training data \cite{jain2020speech}. Equations \ref{eqn3} and \ref{eqn4} shows how SVM works: 

\begin{equation}
\label{eqn3}
   positive equation = w . u + b > 0   
\end{equation}
 
\begin{equation}
\label{eqn4}
   negative equation = w . u + b < 0
\end{equation}

 \begin{itemize}
 \item $ {w} $ = the vector per perpendicular to median of hyper plane
 \item $ {u} $ = the unknown vectors
 \item $ {b} $ = b is constraint
\end{itemize}

The Neural network model is inspired by the human brain \cite {hamid2016ids}. A neural network consists of an input layer, hidden layer and output layer which are all interconnected. The aim is to transform a set of inputs to the desired outputs by using weights associated with the neurons in the hidden layer \cite {hassan2017intrusion}. Neural network passes training input, output is observed. If the output is incorrect then the hidden layer is adjusted until the correct output is achieved. Then the testing data can be passed through the model as the input data and the output is the prediction \cite {biswas2018intrusion}.

    \begin{equation}
    \label{eqn5}
    f\bigg(b + \sum_{i=1}^{n}x_iw_i\bigg)
    \end{equation}
    
    \begin{itemize}
    \item $ {b} $ = bias
    \item $ {x} $ = input to neuron
    \item $ {w} $ = weights
    \item $ {n} $ = the number of inputs from the incoming layer
    \item $ {i} $ = a counter from 0 to n
    \end{itemize}

Two experiments are done using each algorithm on the dataset. The first experiment makes use of 10 fold cross validation. 10 fold cross validation is used to test machine learning models where the data is divided into training and testing data. 10 refer to the number of groups. Each group takes a turn as the test data and the rest of the groups are used as training data. This ensures that there is variance in the test data. The results of the 10 runs are then averaged to give the final results \cite{tandon2019bitcoin}. The second experiment uses the train test split method where the dataset is split 70/30. 70 \% of the dataset is used to train the dataset and 30 \% of the dataset is used for testing. The results of this paper will use the performance metrics of Accuracy, Precision, Recall and F1-score. These performance metrics are calculated by looking at four classification values. The classification values are True Positive (TP), True Negative (TN), False Positive (FP) and False Negative. The equations for how the performance metrics are calculated are shown in equations \ref{eqn6}, \ref{eqn7}, \ref{eqn8} and \ref{eqn9}.

The accuracy displays the total number of correct classifications versus the total classifications made.

    \begin{equation}
    \label{eqn6}
    Accuracy = \frac{TP + TN}{TP + TN + FP + FN}
    \end{equation}

Precision metric is used to measure one of the classifications against how precise it is in comparison to all classifications. The results are presented as an average between both sitting and standing.
\begin{equation}
\label{eqn7}
    Precision = \frac{TP}{TP + FP}
\end{equation}
The recall is used to show the ratio of the correct classification to all classifications for that label. This is run for both sitting and standing and presented as an average.
\begin{equation}
\label{eqn8}
    Recall = \frac{TP}{TP + FN}
\end{equation}
The F1-score is used to provide an average between the Precision and Recall Metrics. 
\begin{equation}
\label{eqn9}
    F1-score = 2 X \frac{Precision * Recall}{Precision + Recall}
\end{equation}

\section{Results and Discussion}
This section presents the output of the machine learning algorithms after they have completed 10 fold cross validation and train test split using the python variables containing the data and comparing the prediction of the data to the actual labels of the data. The performance metrics used to compare the algorithms include the accuracy score as well as precision, recall and f1 score. A confusion matrix is also provided to show how each sample has been classified. 

\subsection{Cross Validation}

\begin{table}[H]
\caption{Cross validation Results} \label{table18}
\centering
\begin{tabular}{||p{3.5cm} p{1.5cm} p{1cm} p{1cm} p{1cm}||}  
 \hline
 Algorithm & Accuracy & Precision & recall & f1-score \\ [0.5ex] 
 \hline\hline
 Random Forest & 92.47 \% &  0.93 & 0.92& 0.92\\ 
 \hline
 K nearest Neighbours & 88.17 \% &   0.89 & 0.88 &0.88 \\
 \hline
 Support Vector Machine & 84.68 \% &   0.86 & 0.85 & 0.85 \\
 \hline
 Neural network model & 90.05 \%&   0.90  &0.90&  0.90 \\  
 \hline
 Ensemble Classifier & 92.18 \%&   0.92  &0.92&  0.92 \\ [1ex] 
 \hline
 \end{tabular}
\end{table}

In table \ref{tab:title1} it can be seen that the best accuracy is from Random Forest followed by the neural network. Although both KNN and Support Vector Machine still have high accuracy. When the algorithms are compiled together in the ensemble classifier the accuracy is 92.18 \%. The accuracy is calculated as an average of the 10 sets of testing data used in each of the 10 cross fold validation process. The dataset is made up of 30 samples each of sitting and standing which each contain 64 subcarriers. So the total number of rows contained in the dataset is 3840 subcarriers. The confusion matrix is a table used to describe how an algorithm has performed. The confusion matrix shows exactly how many samples were classified in which category. The Y axis on the confusion matrix represents the prediction of the algorithm and the X axis represents the actual classification. 

\begin{figure}[H]
    \centering
    \includegraphics[width=0.6\textwidth]{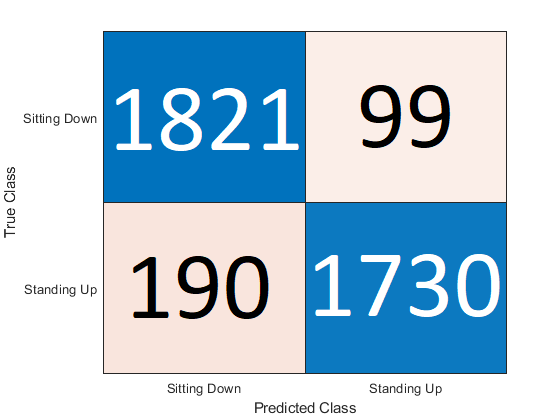}
    \caption{Confusion matrix for Random Forest}
    \label{fig:overview}
\end{figure}

The Random Forest algorithm was the best performer out of all the algorithms. It can be seen on Figure \ref{fig:overview} how the 3840 samples have been classified. 1821 sitting samples were correctly classified as sitting. This is represented in the top left square where the X axis matches the Y axis. Then 99 sitting samples were incorrectly classified as standing. This is where the X axis and Y axis mismatch. The majority of sitting samples were correctly classified so this shows good results. The classification of standing samples was slightly less accurate but still good results. 190 samples were classified incorrectly as sitting, which is higher than the 99 sitting samples incorrectly classified as standing. This leaves the remaining 1730 standing samples as being correctly classified.

\begin{figure}[H]
    \centering
    \includegraphics[width=0.6\textwidth]{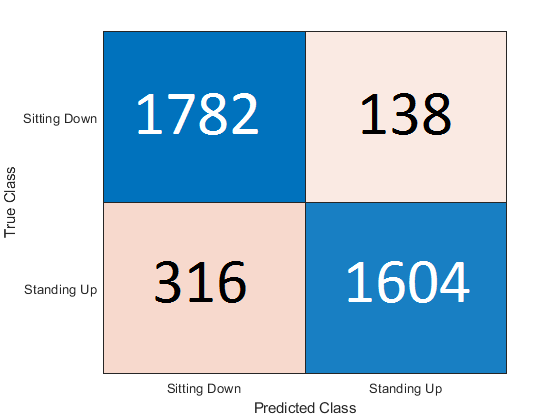}
    \caption{Confusion matrix for KNN}
    \label{fig4}
\end{figure}

The KNN algorithm had an accuracy score of 88.17 \%, which is only around 4 \% less than Random Forest. In the confusion matrix shown in Figure \ref{fig4} it can be observed on how much the classifications differ in the 4 \% difference in accuracy. It appears that both algorithms had better classification results with sitting over standing. KNN had 138 sitting subcarriers incorrectly classified as standing but had 316 standing classifiers incorrectly classified as sitting. However the majority of subcarriers were classified correctly.

\begin{figure}[H]
    \centering
    \includegraphics[width=0.6\textwidth]{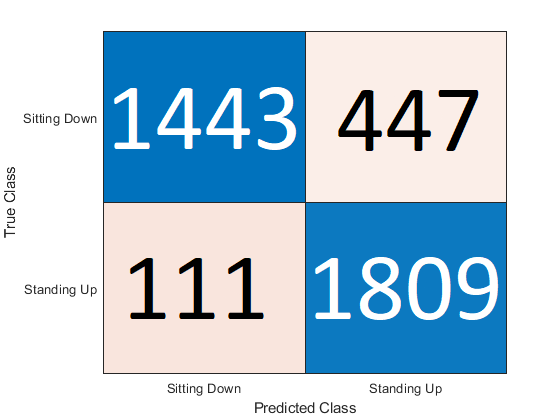}
    \caption{Confusion matrix for SVM}
    \label{fig5}
\end{figure}

Support Vector Machine was the lowest scoring algorithm in this experiment but with an accuracy score of 84.68 \%, the majority of samples were classified correctly. Unlike Random forest and KNN, SVM showed better performance with the standing up samples. Only 111 of the standing subcarriers were wrongly classified as sitting down. 477 sitting down samples were classified incorrectly as standing. As shown in Figure \ref{fig5}. 

\begin{figure}[H]
    \centering
    \includegraphics[width=0.6\textwidth]{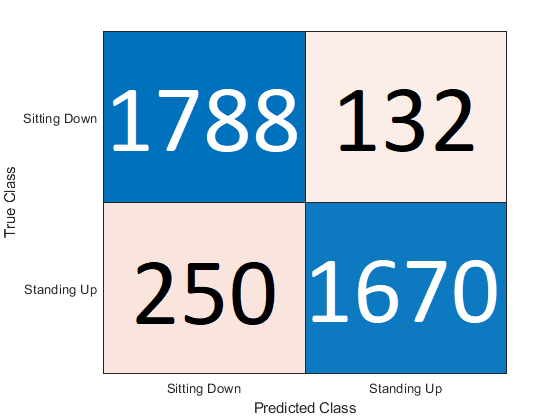}
    \caption{Confusion matrix for Neural Networks}
    \label{fig6}
\end{figure}

The Neural Network classifier had the second best accuracy score of 90.05 \%. Like Random forest and KNN, it had better performance with sitting down samples. The confusion matrix shown in Figure \ref{fig6} shows only 132 sitting samples were incorrectly classified compared to the 250 standing samples classified incorrectly.

\begin{figure}[H]
    \centering
    \includegraphics[width=0.6\textwidth]{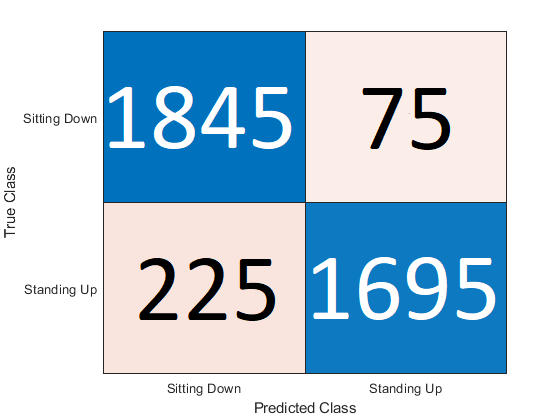}
    \caption{Confusion matrix for Ensemble Classification}
    \label{fig7}
\end{figure}

The confusion matrix for the ensemble classification is shown in Figure \ref{fig7}. The ensemble has the best performance with the sitting down samples with only 75 of the samples being classified as incorrect. The ensemble classifier was let down by the standing up samples as it incorrectly classified 225 samples. It can be seen that the ensemble technique worked well with the sitting down samples but was not so good with the standing up samples. Support Vector Machine had the lowest error rate for standing up samples. 

\subsection{Train Test Split}

\begin{table}[H]
\caption{Train Test Split results} \label{table18}
\centering
\begin{tabular}{||p{4cm} p{1cm} p{1cm} p{1cm} p{1cm}||}  
 \hline
 Algorithm & Accuracy & Precision & recall & f1-score \\ [0.5ex] 
 \hline\hline
 Random Forest & 96.70 \% &  0.97 & 0.97& 0.972\\ 
 \hline
 K nearest Neighbours & 90.71 \% &   0.91 & 0.91 &0.91 \\
 \hline
 Support Vector Machine & 81.77 \% &   0.87 & 0.82 & 0.82 \\
 \hline
 Neural network model & 93.40 \%&   0.94  &0.93&  0.93 \\  
 \hline
 Ensemble Classifier & 93.83 \%&   0.94  &0.94&  0.94 \\ [1ex] 
 \hline
 \end{tabular}
\end{table}

In table \ref{table18} it can be seen that the best accuracy is still Random Forest followed by the neural network. Although both KNN and Support Vector Machine still have high accuracy. When the algorithms are compiled together in the ensemble classifier the accuracy is 93.83 \%. The accuracy is calculated by comparing the 30 \% test data predictions to the actual labels of the data. The full dataset is made up of 30 samples each of sitting and standing which each contain 64 subcarriers. So the total number of rows contained in the dataset is 3840 subcarriers. 1152 subcarriers is the number of the 30 \% test samples used in the train test split method rather than the whole dataset being used testing data at some point. In the testing data there are 512 standing up samples and 640 sitting down samples. The confusion matrix in this experiment shows only the 1152 samples, the total number of tested samples.  

\begin{figure}[H]
    \centering
    \includegraphics[width=0.6\textwidth]{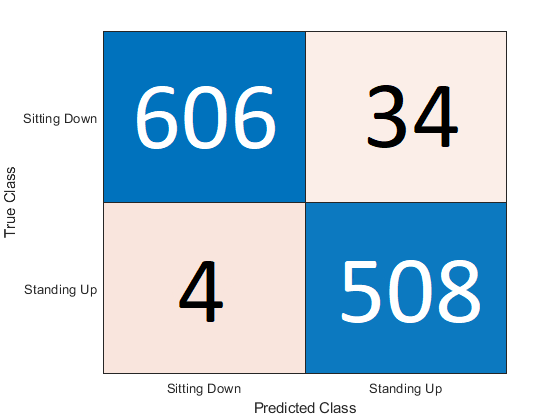}
    \caption{Confusion matrix for Random Forest}
    \label{fig20}
\end{figure}

The Random Forest algorithm was the best performer out of all the algorithms. It can be seen on Figure \ref{fig20} how the 1152 samples have been classified. 606 sitting samples were correctly classified as sitting. This is represented in the top left square where the X axis matches the Y axis. Then 34 sitting samples were incorrectly classified as standing. This is where the X axis and Y axis mismatch. The majority of sitting samples were correctly classified which is a positive result. The classification of standing samples was more accurate than sitting in contrast to the cross validation results. Only 4 samples were classified incorrectly as sitting this leaves the remaining 508 standing samples as being correctly classified.

\begin{figure}[H]
    \centering
    \includegraphics[width=0.6\textwidth]{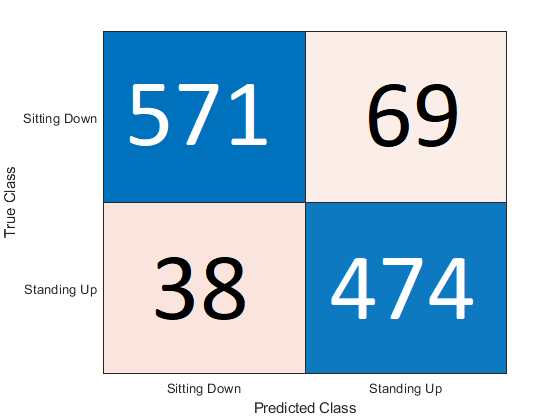}
    \caption{Confusion matrix for KNN}
    \label{fig21}
\end{figure}

The KNN algorithm had an accuracy score of 90.71 \%, which is an improvement over the cross validation experiment. In the confusion matrix shown in Figure \ref{fig21}, KNN just like Random Forest performed better with the standing up samples rather than the sitting down samples. KNN had 69 sitting subcarriers incorrectly classified as standing but had only 38 standing classifiers incorrectly classified as sitting. However the majority of subcarriers were classified correctly.

\begin{figure}[H]
    \centering
    \includegraphics[width=0.6\textwidth]{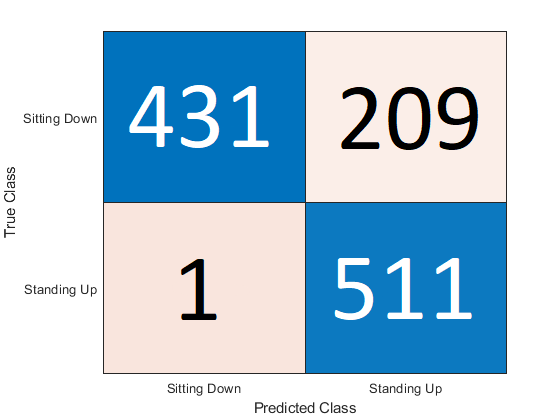}
    \caption{Confusion matrix for SVM}
    \label{fig22}
\end{figure}

Support Vector Machine was the lowest scoring algorithm in this experiment but with an accuracy score of 81.77 \%, the majority of samples were classified correctly. Like Random forest and KNN, SVM showed better performance with the standing up samples. Only 1 of the standing subcarriers was wrongly classified as sitting down however 209 sitting down samples were classified incorrectly as standing, as shown in Figure \ref{fig22}. 

\begin{figure}[H]
    \centering
    \includegraphics[width=0.6\textwidth]{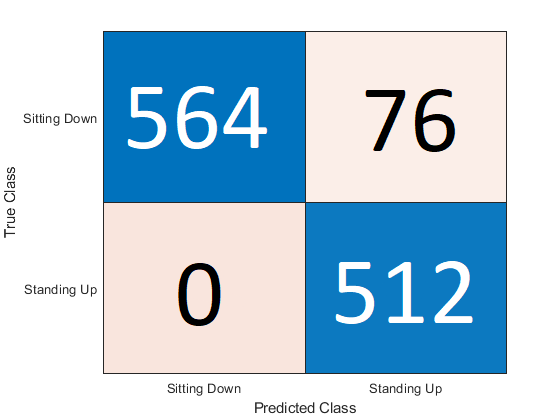}
    \caption{Confusion matrix for Neural Networks}
    \label{fig23}
\end{figure}

The Neural Network classifier had the second best accuracy score of 93.40 \%. Like the other algorithms, it had better performance with standing up samples. The confusion matrix shown in Figure \ref{fig23} shows 76 sitting samples were incorrectly classified compared to the 0 standing samples classified incorrectly.\\
\\
\\

\begin{figure}[H]
    \centering
    \includegraphics[width=0.6\textwidth]{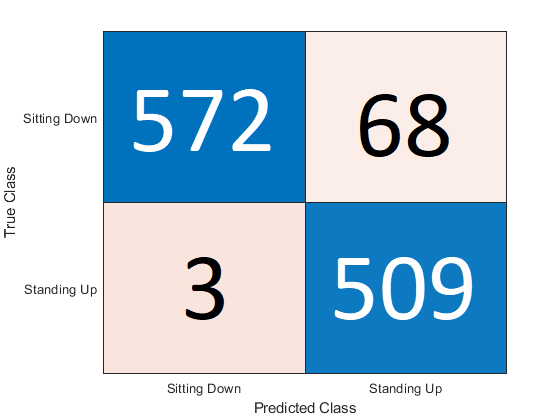}
    \caption{Confusion matrix for Ensemble Classification}
    \label{fig24}
\end{figure}

The confusion matrix for the ensemble classification is shown in Figure \ref{fig24}. The ensemble method shows better performance with the standing samples as expected as all the algorithms performed better with the standing samples. The ensemble method gave a good average number for the incorrect sitting samples preventing it going to high, making use of the voting system. 

\subsection{Comparison of Cross Validation and Train Test Split}
\begin{figure}[H]
    \centering
    \includegraphics[width=0.6\textwidth]{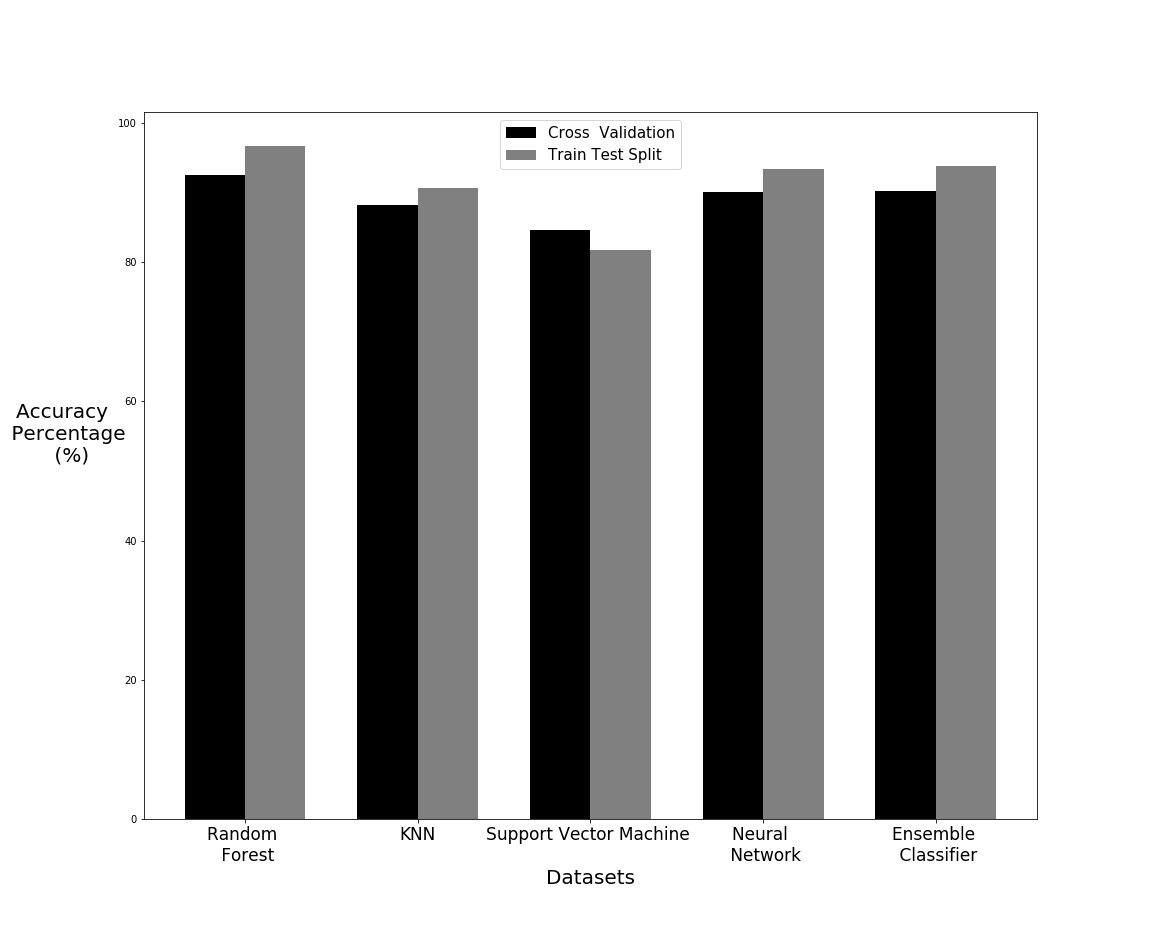}
    \caption{Comparison of Cross Validation and Train Test Split}
    \label{fig25}
\end{figure}

The difference in accuracy can be seen in Figure \ref{fig25}. The train test split shows better classification results with the standing up samples. This is because there are more standing up samples within the 70 \% training set. This shows that the more training on a sample gives better results. All of the algorithms have higher accuracy except from SVM with the train test split. Cross validation however gives a better representation of the algorithm performance since all of the data takes turn of training and testing so every possible combination is tested.

\subsection{Real Time classification}
For Real Time classification of data the dataset needs to be used to create a model. As Random forest provided the highest accuracy results, it was chosen to create the model. Instead of dividing the dataset into 10 groups for cross fold validation, the whole dataset was used for training. This allows for the model to have the most amount of training. The SciKit Python package allows for models to be saved and recalled later by using the Joblib package. Flask was used to create a web interface which could action Python scripts. 

\begin{figure}[H]
    \centering
    \includegraphics[width=0.6\textwidth]{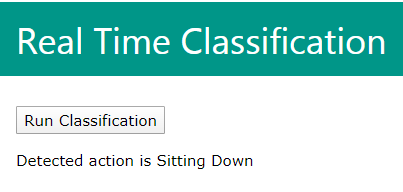}
    \caption{Flask web interface displaying classification result}
    \label{fig10}
\end{figure}

\begin{figure}[H]
    \centering
    \includegraphics[width=0.6\textwidth]{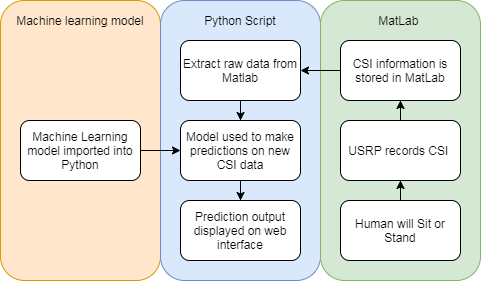}
    \caption{Flask web interface process}
    \label{fig11}
\end{figure}

The application works when the user presses the "Run Classification" button. The button then actions a Python script within the Flask app. The Python script works by connecting to the Matlab session that records the CSI from the USRP. The Matlab session will need to be shared and then Python can connect and access the variables stored on Matlab. When an experiment is ran on the USRP the CSI is stored in a timeseries called CSI in Matlab. The Python script first activates a Matlab script which then extracts the raw CSI data from the timeseries. Once the raw data is stored on a variable in Matlab the Python script can access the variable and apply the previously saved model to make classifications on the new data obtained from the USRPs. As this process take place the interface will display "Loading..." as the output. To test the real time application additional samples of sitting down and standing up were taken. Six of each group were taken to give a total of 12 samples. These 12 samples were completely unseen when training the model as they were not contained in the dataset. The trained model was able to correctly classify all of these samples. As seen in Figure \ref{fig10}, the classification is displayed as the output after the script has run. This web application has proved to be able to access the Matlab variable that contains the CSI obtained from the USRP and make classifications using a previously stored model. The real time web application is able to be extended to make any amount of classifications as it is based on the model used to make the classifications of newly received data. Figure \ref{fig11} details the process undertaken by the real time application web interface.

\subsection{Benchmark dataset}
As the machine learning results for the dataset were of high accuracy, it evidences that CSI is a viable method for human motion detection. For a comparison on how effective CSI can be to identify human motion we have compared the machine learning results of this dataset to that of another dataset. \cite{anguita2013public} have published a dataset detecting a range of human motions using smart phones which are equipped with accelerometers. The machine learning process used with the USRP dataset created in this research has been applied to this benchmark training dataset. This comparison gives a good indication of how non-interference detection compares to wearable devices in the field of human motion detection.

\begin{table}[H]
\caption{Comparison of results with cross validation} \label{table005}
\centering
\begin{tabular}{||p{4cm} p{2cm} p{2cm}||}  
 \hline
 Algorithm & USRP Dataset Accuracy & Benchmark Dataset accuracy\\ [0.5ex] 
 \hline\hline
 Random Forest & 92.47 \% &  91.20 \%\\ 
 \hline
 K nearest Neighbours & 88.17 \% &   77.06 \%\\
 \hline
 Support Vector Machine & 84.68 \% &  85.90 \%\\
 \hline
 Neural network model & 90.05 \%&   89.21 \%\\
 \hline
 Ensemble Classifier & 92.18 \%&   92.40 \%\\ [1ex] 
 \hline
 \end{tabular}
\end{table}

\begin{table}[H]
\caption{Comparison of results with train test split} \label{table006}
\centering
\begin{tabular}{||p{4cm} p{2cm} p{2cm}||}  
 \hline
 Algorithm & USRP Dataset Accuracy & Benchmark Dataset accuracy\\ [0.5ex] 
 \hline\hline
 Random Forest & 96.70 \% &  96.49 \%\\ 
 \hline
 K nearest Neighbours & 90.71 \% &   92.48 \%\\
 \hline
 Support Vector Machine & 81.77 \% &  86.21 \%\\
 \hline
 Neural network model & 93.40 \%&   96.11 \%\\
 \hline
 Ensemble Classifier & 93.83 \%&   97.74 \%\\ [1ex] 
 \hline
 \end{tabular}
\end{table}

\begin{figure}[H]
    \centering
    \includegraphics[width=0.6\textwidth]{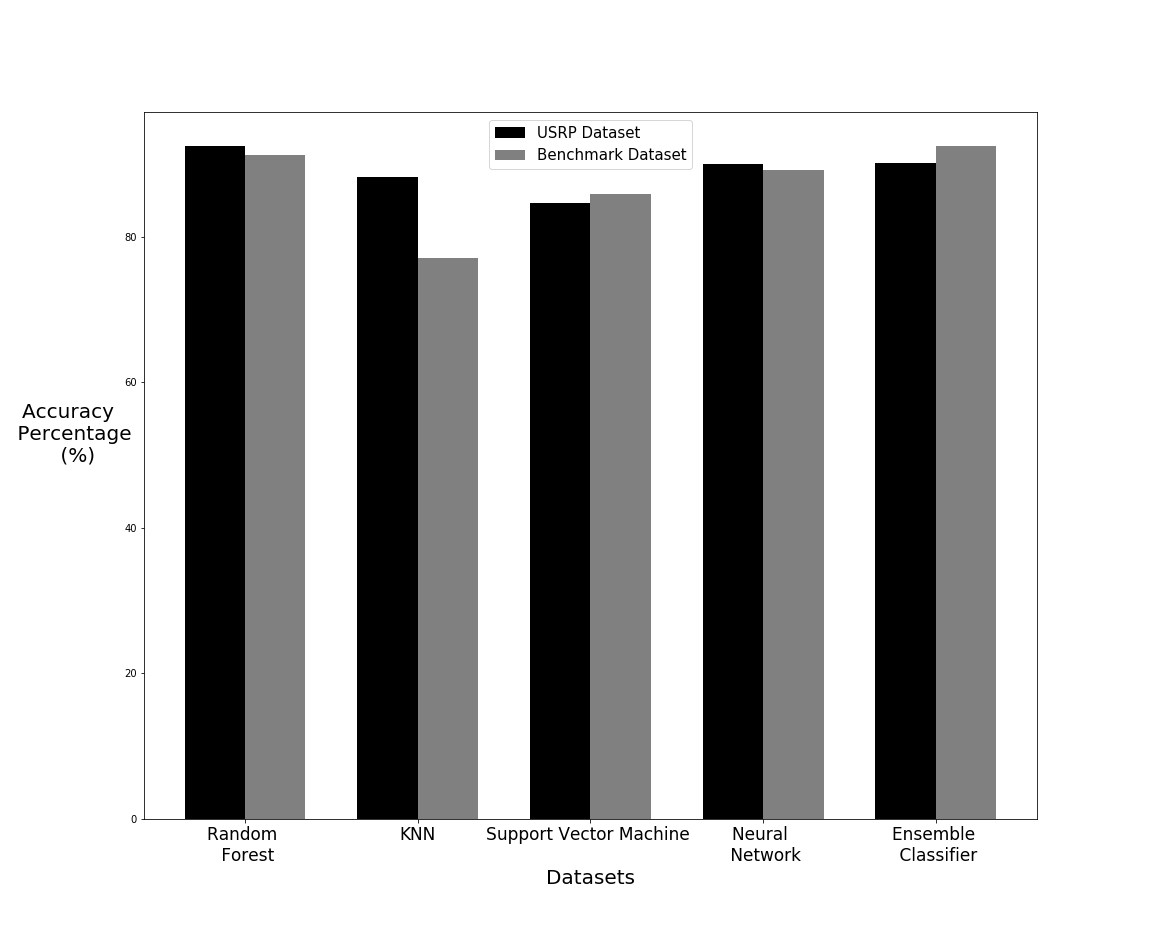}
    \caption{Comparison of results with cross validation}
    \label{fig1201}
\end{figure}

\begin{figure}[H]
    \centering
    \includegraphics[width=0.6\textwidth]{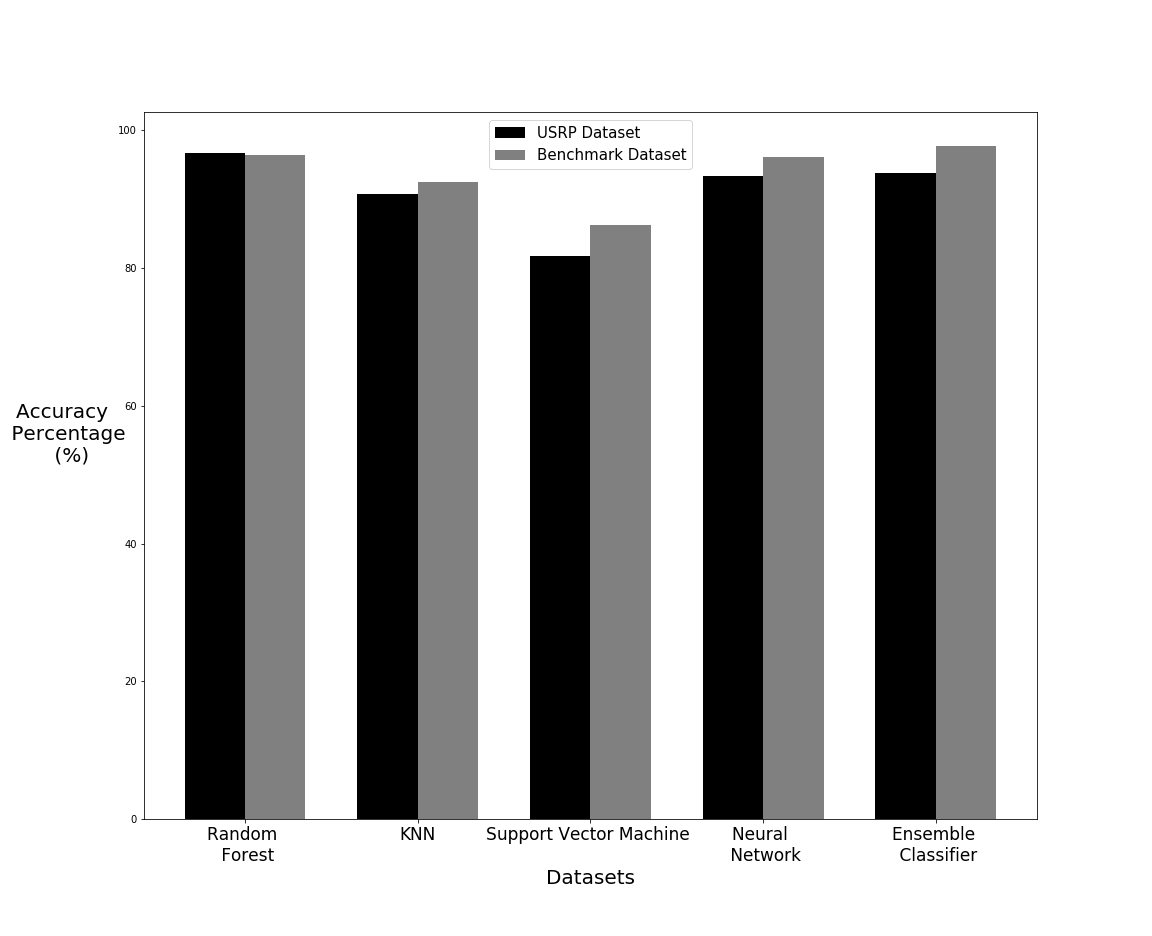}
    \caption{Comparison of results with train test split}
    \label{fig12}
\end{figure}

The results show that the USRP dataset is able to provide similar results to the benchmark dataset which is using wearable devices. The Random Forest algorithm displays similar results. The accuracy values are shown in tables \ref{table005} and \ref{table006} for cross validation and train test split experiments respectfully. Figures \ref{fig1201} and \ref{fig12} give visual representation of the differences between the two datasets for cross validation and train test split experiments respectfully. The Random Forest was the best performer in both sets of data with both cross validation and train test split methods. KNN performed much better using the USRP dataset with a cross validation but was lower with train test split. Support Vector Machine had similar performance within the two datasets with only a larger difference in accuracy between datasets using the train split method. The Neural Network algorithm also had a small difference between datasets with a slight increase with the cross validated USRP dataset but a larger difference in favour of the benchmark dataset when using train test split. The ensemble classifier actually performed better with the benchmark dataset in both methods but by only a small difference when using the cross validation method. Such findings demonstrate that the USRP is capable of producing similar results and even higher precision scores compared to a dataset obtained using wearable devices.The primary reason that the datasets collected using USRP outperforms the wearable devices datasets is that USRP leveage on multiple frequency subcarriers. An intricate change in wireless medium is picked up by the multiple carrier USRP model, whereas the wearable devices such as accelerometer and magnetometer as not sensitive enough against body motion. That is why, due to high sensitivity against body motion, the USRP works better in detecting body movements.

\section{Conclusion}
In this paper we have proposed an algorithm and dataset which can be used in the detection of human motion. The dataset includes observations of the channel state information from USRPs as human activities takes place between the antennas. The dataset is designed for binary classification between sitting down and standing up human motion. The performance of machine learning  show good results with the Random Forest algorithm producing a high accuracy result of 92.47 \%. The high accuracy in the results show that there is a significant difference between the CSI information of standing up and sitting down for a machine algorithm to be able to establish the difference. The web application was able to successfully classify samples of data that were absent during the learning phase directly from the Matlab session which contained the CSI directly from the USRP. The use of USRP data to detect human motion was also compared to a benchmark dataset where human motion was detected using wearable devices. The same machine learning techniques were applied to the benchmark dataset and the results shows good accuracy with the benchmark dataset. 

\section{Acknowledgement}
William Taylor’s studentship is funded by CENSIS UK through Scottish funding council in collaboration with British Telecom. This work is supported in parts by EPSRC DTG EP/N509668/1 Eng, EP/T021020/1 and EP/T021063/1


\reftitle{References}
\bibliography{mdpi}

\begin{thebibliography}{-------}
\providecommand{\natexlab}[1]{#1}

\bibitem[Yang \em{et~al.}(2020)Yang, Ren, Chen, Wang, and Ding]{yang2020human}
Yang, X.; Ren, X.; Chen, M.; Wang, L.; Ding, Y.
\newblock Human Posture Recognition in Intelligent Healthcare.
\newblock  Journal of Physics: Conference Series. IOP Publishing,  2020, Vol.
  1437, p. 012014.

\bibitem[Abbasi \em{et~al.}(2016)Abbasi, Rehman, Qaraqe, and
  Alomainy]{abbasi2016advances}
Abbasi, Q.H.; Rehman, M.U.; Qaraqe, K.; Alomainy, A.
\newblock {\em Advances in body-centric wireless communication: Applications
  and state-of-the-art}; Institution of Engineering and Technology,  2016.

\bibitem[Dong \em{et~al.}(2017)Dong, Ren, Shah, Hu, Zhao, Yang, Haider, Zhang,
  Zhao, and Abbasi]{dong2017monitoring}
Dong, B.; Ren, A.; Shah, S.A.; Hu, F.; Zhao, N.; Yang, X.; Haider, D.; Zhang,
  Z.; Zhao, W.; Abbasi, Q.H.
\newblock Monitoring of atopic dermatitis using leaky coaxial cable.
\newblock {\em Healthcare technology letters} {\bf 2017}, {\em 4},~244--248.

\bibitem[Al-Mishmish \em{et~al.}(2018)Al-Mishmish, Alkhayyat, Rahim, Hammood,
  Ahmad, and Abbasi]{al2018critical}
Al-Mishmish, H.; Alkhayyat, A.; Rahim, H.A.; Hammood, D.A.; Ahmad, R.B.;
  Abbasi, Q.H.
\newblock Critical data-based incremental cooperative communication for
  wireless body area network.
\newblock {\em Sensors} {\bf 2018}, {\em 18},~3661.

\bibitem[Mercuri \em{et~al.}(2016)Mercuri, Garripoli, Karsmakers, Soh,
  Vandenbosch, Pace, Leroux, and Schreurs]{mercuri2016healthcare}
Mercuri, M.; Garripoli, C.; Karsmakers, P.; Soh, P.J.; Vandenbosch, G.A.; Pace,
  C.; Leroux, P.; Schreurs, D.
\newblock Healthcare system for non-invasive fall detection in indoor
  environment. In {\em Applications in Electronics Pervading Industry,
  Environment and Society}; Springer,  2016; pp. 145--152.

\bibitem[Haider \em{et~al.}(2019)Haider, Ren, Fan, Zhao, Yang, Shah, Hu, and
  Abbasi]{haider2019efficient}
Haider, D.; Ren, A.; Fan, D.; Zhao, N.; Yang, X.; Shah, S.A.; Hu, F.; Abbasi,
  Q.H.
\newblock An efficient monitoring of eclamptic seizures in wireless sensors
  networks.
\newblock {\em Computers \& Electrical Engineering} {\bf 2019}, {\em
  75},~16--30.

\bibitem[Liu \em{et~al.}(2019)Liu, Zhang, Yang, Zhou, Ren, Wang, Liu, Pang, and
  Deen]{liu2019novel}
Liu, Y.; Zhang, L.; Yang, Y.; Zhou, L.; Ren, L.; Wang, F.; Liu, R.; Pang, Z.;
  Deen, M.J.
\newblock A novel cloud-based framework for the elderly healthcare services
  using digital twin.
\newblock {\em IEEE Access} {\bf 2019}, {\em 7},~49088--49101.

\bibitem[Fan \em{et~al.}(2018)Fan, Ren, Zhao, Yang, Zhang, Shah, Hu, and
  Abbasi]{fan2018breathing}
Fan, D.; Ren, A.; Zhao, N.; Yang, X.; Zhang, Z.; Shah, S.A.; Hu, F.; Abbasi,
  Q.H.
\newblock Breathing rhythm analysis in body centric networks.
\newblock {\em IEEE Access} {\bf 2018}, {\em 6},~32507--32513.

\bibitem[Shang \em{et~al.}(2019)Shang, Chang, Chen, Zhao, and
  Chen]{shang2019bia}
Shang, C.; Chang, C.Y.; Chen, G.; Zhao, S.; Chen, H.
\newblock BIA: Behavior Identification Algorithm using Unsupervised Learning
  Based on Sensor Data for Home Elderly.
\newblock {\em IEEE Journal of Biomedical and Health Informatics} {\bf 2019}.

\bibitem[Yang \em{et~al.}(2018)Yang, Shah, Ren, Zhao, Zhao, Hu, Zhang, Zhao,
  Rehman, and Alomainy]{yang2018monitoring}
Yang, X.; Shah, S.A.; Ren, A.; Zhao, N.; Zhao, J.; Hu, F.; Zhang, Z.; Zhao, W.;
  Rehman, M.U.; Alomainy, A.
\newblock Monitoring of patients suffering from REM sleep behavior disorder.
\newblock {\em IEEE Journal of Electromagnetics, RF and Microwaves in Medicine
  and Biology} {\bf 2018}, {\em 2},~138--143.

\bibitem[Zhang \em{et~al.}(2019)Zhang, Song, Chen, Zhang, and
  Zhuang]{zhang2019wigrus}
Zhang, T.; Song, T.; Chen, D.; Zhang, T.; Zhuang, J.
\newblock WiGrus: A Wifi-Based Gesture Recognition System Using
  Software-Defined Radio.
\newblock {\em IEEE Access} {\bf 2019}, {\em 7},~131102--131113.

\bibitem[Tahir \em{et~al.}(2019)Tahir, Ahmad, Shah, Morison, Skelton, Larijani,
  Abbasi, Imran, and Gibson]{tahir2019wifreeze}
Tahir, A.; Ahmad, J.; Shah, S.A.; Morison, G.; Skelton, D.A.; Larijani, H.;
  Abbasi, Q.H.; Imran, M.A.; Gibson, R.M.
\newblock WiFreeze: Multiresolution Scalograms for Freezing of Gait Detection
  in Parkinsons Leveraging 5G Spectrum with Deep Learning.
\newblock {\em Electronics} {\bf 2019}, {\em 8},~1433.

\bibitem[Liu \em{et~al.}(2018)Liu, Shah, Zhao, and Yang]{liu2018respiration}
Liu, L.; Shah, S.A.; Zhao, G.; Yang, X.
\newblock Respiration symptoms monitoring in body area networks.
\newblock {\em Applied Sciences} {\bf 2018}, {\em 8},~568.

\bibitem[Yang \em{et~al.}(2020)Yang, Fan, Ren, Zhao, Shah, Alomainy, Ur-Rehman,
  and Abbasi]{yang2020diagnosis}
Yang, X.; Fan, D.; Ren, A.; Zhao, N.; Shah, S.A.; Alomainy, A.; Ur-Rehman, M.;
  Abbasi, Q.H.
\newblock Diagnosis of the Hypopnea syndrome in the early stage.
\newblock {\em Neural Computing and Applications} {\bf 2020}, {\em
  32},~855--866.

\bibitem[Demir \em{et~al.}(2016)Demir, Abbasi, Ankarali, Alomainy, Qaraqe,
  Serpedin, and Arslan]{demir2016anatomical}
Demir, A.F.; Abbasi, Q.H.; Ankarali, Z.E.; Alomainy, A.; Qaraqe, K.; Serpedin,
  E.; Arslan, H.
\newblock Anatomical region-specific in vivo wireless communication channel
  characterization.
\newblock {\em IEEE journal of biomedical and health informatics} {\bf 2016},
  {\em 21},~1254--1262.

\bibitem[Santos \em{et~al.}(2019)Santos, Endo, Monteiro, Rocha, Silva, and
  Lynn]{santos2019accelerometer}
Santos, G.L.; Endo, P.T.; Monteiro, K.H.d.C.; Rocha, E.d.S.; Silva, I.; Lynn,
  T.
\newblock Accelerometer-based human fall detection using convolutional neural
  networks.
\newblock {\em Sensors} {\bf 2019}, {\em 19},~1644.

\bibitem[Jilani \em{et~al.}(2018)Jilani, Munoz, Abbasi, and
  Alomainy]{jilani2018millimeter}
Jilani, S.F.; Munoz, M.O.; Abbasi, Q.H.; Alomainy, A.
\newblock Millimeter-wave liquid crystal polymer based conformal antenna array
  for 5G applications.
\newblock {\em IEEE Antennas and Wireless Propagation Letters} {\bf 2018}, {\em
  18},~84--88.

\bibitem[Yao \em{et~al.}(2019)Yao, Khan, and Jin]{yao2019energy}
Yao, X.; Khan, A.; Jin, Y.
\newblock Energy Efficient Communication among Wearable Devices using Optimized
  Motion Detection.
\newblock  2019 IEEE Symposium on Computers and Communications (ISCC). IEEE,
  2019, pp. 1--6.

\bibitem[Yang \em{et~al.}(2011)Yang, Abbasi, Alomainy, and
  Hao]{yang2011spatial}
Yang, X.D.; Abbasi, Q.H.; Alomainy, A.; Hao, Y.
\newblock Spatial correlation analysis of on-body radio channels considering
  statistical significance.
\newblock {\em IEEE Antennas and Wireless Propagation Letters} {\bf 2011}, {\em
  10},~780--783.

\bibitem[Zhao \em{et~al.}(2019)Zhao, Liu, Wei, Zhang, Wang, and Fan]{zhao2019r}
Zhao, J.; Liu, L.; Wei, Z.; Zhang, C.; Wang, W.; Fan, Y.
\newblock R-DEHM: CSI-based robust duration estimation of human motion with
  WiFi.
\newblock {\em Sensors} {\bf 2019}, {\em 19},~1421.

\bibitem[Chopra \em{et~al.}(2016)Chopra, Yang, Abbasi, Qaraqe, Philpott, and
  Alomainy]{chopra2016thz}
Chopra, N.; Yang, K.; Abbasi, Q.H.; Qaraqe, K.A.; Philpott, M.; Alomainy, A.
\newblock THz time-domain spectroscopy of human skin tissue for in-body
  nanonetworks.
\newblock {\em IEEE Transactions on Terahertz Science and Technology} {\bf
  2016}, {\em 6},~803--809.

\bibitem[Lolla and Zhao(2019)]{lolla2019wifi}
Lolla, S.; Zhao, A.
\newblock WiFi Motion Detection: A Study into Efficacy and Classification.
\newblock  2019 IEEE Integrated STEM Education Conference (ISEC). IEEE,  2019,
  pp. 375--378.

\bibitem[Christiansen and Smith(2019)]{christiansen2019development}
Christiansen, J.M.; Smith, G.E.
\newblock Development and Calibration of a Low-Cost Radar Testbed Based on the
  Universal Software Radio Peripheral.
\newblock {\em IEEE Aerospace and Electronic Systems Magazine} {\bf 2019}, {\em
  34},~50--60.

\bibitem[Kim(2017)]{kim2017device}
Kim, S.C.
\newblock Device-free activity recognition using CSI \& big data analysis: A
  survey.
\newblock  2017 Ninth International Conference on Ubiquitous and Future
  Networks (ICUFN). IEEE,  2017, pp. 539--541.

\bibitem[Tichy and Ulovec(2012)]{tichy2012ofdm}
Tichy, M.; Ulovec, K.
\newblock OFDM system implementation using a USRP unit for testing purposes.
\newblock  Proceedings of 22nd International Conference Radioelektronika 2012.
  IEEE,  2012, pp. 1--4.

\bibitem[Ashleibta \em{et~al.}(2020)Ashleibta, Shah, Zahid, Imran, and
  Abbasi]{ashleibta2020software}
Ashleibta, A.; Shah, S.; Zahid, A.; Imran, M.A.; Abbasi, Q.H.
\newblock Software Defined Radio Based Testbed for Large Scale Body Movements
  {\bf 2020}.

\bibitem[Zhang \em{et~al.}(2017)Zhang, Xu, Hu, and Kanhere]{zhang2017wicare}
Zhang, J.; Xu, W.; Hu, W.; Kanhere, S.S.
\newblock WiCare: Towards In-Situ Breath Monitoring.
\newblock  Proceedings of the 14th EAI International Conference on Mobile and
  Ubiquitous Systems: Computing, Networking and Services,  2017, pp. 126--135.

\bibitem[Chin \em{et~al.}(2019)Chin, Ng, Yap, Tong, Ho, and Goh]{chin2019daily}
Chin, Z.H.; Ng, H.; Yap, T.T.V.; Tong, H.L.; Ho, C.C.; Goh, V.T.
\newblock Daily Activities Classification on Human Motion Primitives Detection
  Dataset. In {\em Computational Science and Technology}; Springer,  2019; pp.
  117--125.

\bibitem[Shah \em{et~al.}(2018)Shah, Fan, Ren, Zhao, Yang, and
  Tanoli]{shah2018seizure}
Shah, S.A.; Fan, D.; Ren, A.; Zhao, N.; Yang, X.; Tanoli, S.A.K.
\newblock Seizure episodes detection via smart medical sensing system.
\newblock {\em Journal of Ambient Intelligence and Humanized Computing} {\bf
  2018}, pp. 1--13.

\bibitem[Fioranelli \em{et~al.}(2019)Fioranelli, Le~Kernec, and
  Shah]{fioranelli2019radar}
Fioranelli, F.; Le~Kernec, J.; Shah, S.A.
\newblock Radar for Health Care: Recognizing Human Activities and Monitoring
  Vital Signs.
\newblock {\em IEEE Potentials} {\bf 2019}, {\em 38},~16--23.

\bibitem[Ding \em{et~al.}(2019)Ding, Zou, Sun, Hong, Zhu, and Li]{ding2019fall}
Ding, C.; Zou, Y.; Sun, L.; Hong, H.; Zhu, X.; Li, C.
\newblock Fall detection with multi-domain features by a portable FMCW radar.
\newblock  2019 IEEE MTT-S International Wireless Symposium (IWS). IEEE,  2019,
  pp. 1--3.

\bibitem[Shah and Fioranelli(2019{\natexlab{a}})]{shah2019human}
Shah, S.A.; Fioranelli, F.
\newblock Human Activity Recognition: Preliminary Results for Dataset
  Portability using FMCW Radar {\bf 2019}.

\bibitem[Shah and Fioranelli(2019{\natexlab{b}})]{shah2019rf}
Shah, S.A.; Fioranelli, F.
\newblock RF sensing technologies for assisted daily living in healthcare: A
  comprehensive review.
\newblock {\em IEEE Aerospace and Electronic Systems Magazine} {\bf 2019}, {\em
  34},~26--44.

\bibitem[Liu and Zhang(2019)]{liu2019noma}
Liu, X.; Zhang, X.
\newblock NOMA-based Resource Allocation for Cluster-based Cognitive Industrial
  Internet of Things.
\newblock {\em IEEE Transactions on Industrial Informatics} {\bf 2019}.

\bibitem[Liu \em{et~al.}(2018)Liu, Jia, Zhang, and Lu]{liu2018novel}
Liu, X.; Jia, M.; Zhang, X.; Lu, W.
\newblock A novel multichannel Internet of things based on dynamic spectrum
  sharing in 5G communication.
\newblock {\em IEEE Internet of Things Journal} {\bf 2018}, {\em
  6},~5962--5970.

\bibitem[Jalal \em{et~al.}(2019)Jalal, Quaid, and Kim]{jalal2019wrist}
Jalal, A.; Quaid, M.A.K.; Kim, K.
\newblock A wrist worn acceleration based human motion analysis and
  classification for ambient smart home system.
\newblock {\em Journal of Electrical Engineering \& Technology} {\bf 2019},
  {\em 14},~1733--1739.

\bibitem[Zhang \em{et~al.}(2019)Zhang, Fu, and Shu]{zhang2019recognizing}
Zhang, H.; Fu, Z.; Shu, K.I.
\newblock Recognizing Ping-Pong Motions Using Inertial Data Based on Machine
  Learning Classification Algorithms.
\newblock {\em IEEE Access} {\bf 2019}, {\em 7},~167055--167064.

\bibitem[Zhang \em{et~al.}(2020)Zhang, Su, Dong, and
  Pahlavan]{zhang2020complex}
Zhang, P.; Su, Z.; Dong, Z.; Pahlavan, K.
\newblock Complex Motion Detection Based on Channel State Information and
  LSTM-RNN.
\newblock  2020 10th Annual Computing and Communication Workshop and Conference
  (CCWC). IEEE,  2020, pp. 0756--0760.

\bibitem[Al-Khafajiy \em{et~al.}()Al-Khafajiy, Otoum, Baker, Asim, Maamar,
  Aloqaily, Taylor, and Randles]{alintelligent}
Al-Khafajiy, M.; Otoum, S.; Baker, T.; Asim, M.; Maamar, Z.; Aloqaily, M.;
  Taylor, M.; Randles, M.
\newblock Intelligent Control and Security of Fog Resources in Healthcare
  Systems via a Cognitive Fog Model.
\newblock {\em ACM Transactions on Internet Technology}.

\bibitem[Oueida \em{et~al.}(2018)Oueida, Kotb, Aloqaily, Jararweh, and
  Baker]{oueida2018edge}
Oueida, S.; Kotb, Y.; Aloqaily, M.; Jararweh, Y.; Baker, T.
\newblock An edge computing based smart healthcare framework for resource
  management.
\newblock {\em Sensors} {\bf 2018}, {\em 18},~4307.

\bibitem[Anjomshoa \em{et~al.}(2017)Anjomshoa, Aloqaily, Kantarci,
  Erol-Kantarci, and Schuckers]{anjomshoa2017social}
Anjomshoa, F.; Aloqaily, M.; Kantarci, B.; Erol-Kantarci, M.; Schuckers, S.
\newblock Social behaviometrics for personalized devices in the internet of
  things era.
\newblock {\em IEEE Access} {\bf 2017}, {\em 5},~12199--12213.

\bibitem[Nipu \em{et~al.}(2018)Nipu, Talukder, Islam, and
  Chakrabarty]{nipu2018human}
Nipu, M.N.A.; Talukder, S.; Islam, M.S.; Chakrabarty, A.
\newblock Human identification using wifi signal.
\newblock  2018 Joint 7th International Conference on Informatics, Electronics
  \& Vision (ICIEV) and 2018 2nd International Conference on Imaging, Vision \&
  Pattern Recognition (icIVPR). IEEE,  2018, pp. 300--304.

\bibitem[Tanoli \em{et~al.}(2018)Tanoli, Rehman, Khan, Jadoon, Ali~Khan, Nawaz,
  Shah, Yang, and Nasir]{tanoli2018experimental}
Tanoli, S.A.K.; Rehman, M.; Khan, M.B.; Jadoon, I.; Ali~Khan, F.; Nawaz, F.;
  Shah, S.A.; Yang, X.; Nasir, A.A.
\newblock An experimental channel capacity analysis of cooperative networks
  using Universal Software Radio Peripheral (USRP).
\newblock {\em Sustainability} {\bf 2018}, {\em 10},~1983.

\bibitem[Hao and Ho(2019)]{hao2019machine}
Hao, J.; Ho, T.K.
\newblock Machine Learning Made Easy: A Review of Scikit-learn Package in
  Python Programming Language.
\newblock {\em Journal of Educational and Behavioral Statistics} {\bf 2019},
  {\em 44},~348--361.

\bibitem[Pappalardo(2019)]{pappalardo2019scikit}
Pappalardo, L.
\newblock SCIKIT-MOBILITY: A PYTHON LIBRARY FOR THE ANALYSIS, GENERATION AND
  RISK ASSESSMENT OF MOBILITY DATA.
\newblock {\em arXiv preprint arXiv:1907.07062} {\bf 2019}.

\bibitem[Shaikhina \em{et~al.}(2019)Shaikhina, Lowe, Daga, Briggs, Higgins, and
  Khovanova]{shaikhina2019decision}
Shaikhina, T.; Lowe, D.; Daga, S.; Briggs, D.; Higgins, R.; Khovanova, N.
\newblock Decision tree and random forest models for outcome prediction in
  antibody incompatible kidney transplantation.
\newblock {\em Biomedical Signal Processing and Control} {\bf 2019}, {\em
  52},~456--462.

\bibitem[Sa{\c{c}}l{\i} \em{et~al.}(2019)Sa{\c{c}}l{\i}, Ayd{\i}nalp,
  Cans{\i}z, Joof, Yilmaz, {\c{C}}ay{\"o}ren, {\"O}nal, and
  Akduman]{saccli2019microwave}
Sa{\c{c}}l{\i}, B.; Ayd{\i}nalp, C.; Cans{\i}z, G.; Joof, S.; Yilmaz, T.;
  {\c{C}}ay{\"o}ren, M.; {\"O}nal, B.; Akduman, I.
\newblock Microwave dielectric property based classification of renal calculi:
  Application of a kNN algorithm.
\newblock {\em Computers in biology and medicine} {\bf 2019}, {\em
  112},~103366.

\bibitem[Li \em{et~al.}(2019)Li, Gu, Zhang, An, and Li]{li2019research}
Li, K.; Gu, Y.; Zhang, P.; An, W.; Li, W.
\newblock Research on KNN Algorithm in Malicious PDF Files Classification under
  Adversarial Environment.
\newblock  Proceedings of the 2019 4th International Conference on Big Data and
  Computing,  2019, pp. 156--159.

\bibitem[Jain \em{et~al.}(2020)Jain, Narayan, Balaji, Bhowmick, Muthu,
  et~al.]{jain2020speech}
Jain, M.; Narayan, S.; Balaji, P.; Bhowmick, A.; Muthu, R.K.; others.
\newblock Speech emotion recognition using support vector machine.
\newblock {\em arXiv preprint arXiv:2002.07590} {\bf 2020}.

\bibitem[Hamid \em{et~al.}(2016)Hamid, Sugumaran, and
  Balasaraswathi]{hamid2016ids}
Hamid, Y.; Sugumaran, M.; Balasaraswathi, V.
\newblock Ids using machine learning-current state of art and future
  directions.
\newblock {\em Current Journal of Applied Science and Technology} {\bf 2016},
  pp. 1--22.

\bibitem[Hassan \em{et~al.}(2017)Hassan, Sheta, and Wahbi]{hassan2017intrusion}
Hassan, A.A.; Sheta, A.F.; Wahbi, T.M.
\newblock Intrusion Detection System Using Weka Data Mining Tool.
\newblock {\em Internaitonal Journal of Science and Research (IJSR)} {\bf
  2017}, {\em 6},~2319--7064.

\bibitem[Biswas(2018)]{biswas2018intrusion}
Biswas, S.K.
\newblock Intrusion detection using machine learning: A comparison study.
\newblock {\em International Journal of Pure and Applied Mathematics} {\bf
  2018}, {\em 118},~101--114.

\bibitem[Tandon \em{et~al.}(2019)Tandon, Tripathi, Saraswat, and
  Dabas]{tandon2019bitcoin}
Tandon, S.; Tripathi, S.; Saraswat, P.; Dabas, C.
\newblock Bitcoin Price Forecasting using LSTM and 10-Fold Cross validation.
\newblock  2019 International Conference on Signal Processing and Communication
  (ICSC). IEEE,  2019, pp. 323--328.

\bibitem[Anguita \em{et~al.}(2013)Anguita, Ghio, Oneto, Parra, and
  Reyes-Ortiz]{anguita2013public}
Anguita, D.; Ghio, A.; Oneto, L.; Parra, X.; Reyes-Ortiz, J.L.
\newblock A public domain dataset for human activity recognition using
  smartphones.
\newblock  Esann,  2013.

\end{thebibliography}

\end{document}